\documentclass[12pt, epsf]{article} 
\input epsf.tex 
\def\be{\begin{equation}} \def\ee{\end{equation}}
\def\bi{\begin{itemize}} \def\ei{\end{itemize}}
\def\bea{\begin{eqnarray}} \def\eea{\end{eqnarray}} \def\ba{\begin{array}}
\def\ea{\end{array}} \def\ben{\begin{enumerate}} \def\een{\end{enumerate}}

\newcommand{\eqn}[1]{(\ref{#1})}

\newcommand{\prl}[3]{Phys. Rev. Lett. {\bf#1} ({#2}) {#3}}

\newcommand{\prd}[3]{Phys. Rev. {\bf D#1} ({#2}) {#3}}
\newcommand{\hepth}[1]{{\tt arXiv:{#1}[hep-th]}}

\def\br{\nonumber\\}

\def\Tr{{\rm Tr}}

\begin{document}
{}~
\hfill \vbox{
\hbox{\today}}\break

\vskip 3.5cm
\centerline{\Large \bf
 AdS Bubbles,
  E$p$-branes and Entanglement}

\vskip 1cm

\vspace*{.5cm}

\centerline{\bf  Harvendra Singh}

\vspace*{.5cm}
\centerline{ \it  Theory Division, Saha Institute of Nuclear Physics} 
\centerline{ \it  1/AF Bidhannagar, Kolkata 700064, India}
\vspace*{.25cm}

\vspace*{.5cm}

\vskip.5cm

\vskip1cm

\centerline{\bf Abstract} \bigskip

The AdS-bubble  solutions interestingly  
 mimic Schr\"odinger-like  geometries
when expressed in light-cone coordinates. 
These D$p$ bubble vacuas exhibit  asymmetric scaling property
with a negative dynamical exponent of time $a<0$, but are
 smooth geometries. Through a time-like T-duality we map these
vacua  to  E$p$-brane bubbles with $a>0$
in type-II$^\ast$  super-strings.
We obtain an expression for
the entanglement entropy for `bubble E3-branes'. It is argued that the
entropy from E3-bubbles  has to be the lowest.     

\vfill 
\eject

\baselineskip=16.2pt


\section{Introduction}

A remarkable progress  has been made
towards understanding  various string backgrounds  which exhibit
Lifshitz or Schr\"odinger type non-relativistic symmetries
\cite{RT}-\cite{Singh:2013}. 
Particularly, in these solutions 
the time and space coordinates  
scale {\it asymmetrically} and therefore
Lorentzian symmetry is explicitly broken in the holographic boundary theory.  
The  Lifshitz type solutions may admit boundary field theories which can
exhibit non-fermi liquid or a strange metallic 
 behavior at ultra-low temperatures. 
The strange metallic effects have also been associated with the 
phenomenon where strongly correlated quantum 
systems develop hidden fermi surfaces \cite{takaya11, subir11}. Recently
estimating the entanglement entropy of  quantum systems
has become an effective holographic tool in order to understand 
strongly coupled CFT dynamics  \cite{RT, takaya11}. 
On general grounds, there are several fundamental issues attached with 
the measurement of the entanglement within the quantum systems, 
including black-holes  \cite{Maldacena:2013xja},
and entangled  pairs \cite{Faulkner:2013ana}. Even for a pure system, once
 it is subdivided into smaller subsystems, say $A$ and $B$,  the subsystems 
get maximally entangled
amongst themselves. The von Neumann entropy measure of the system $A$,   
$ S_A\sim -\Tr\, \rho_A \ln\rho_A$, is defined in terms of 
the reduced density matrix $\rho_A=\Tr _B [\rho]$.  
Following AdS/CFT holographic dictionary the
basic picture is well laid out. That is, if  strongly correlated 
field theory system at  critical point could be represented as a
system living on the boundary of some known bulk AdS theory, 
the boundary system becomes  phenomenologically  more tractable. According 
to Ryu-Takayanagi proposal, in such 
holographic cases  the entanglement
entropy  of the boundary (quantum) theory can also be estimated geometrically 
as the  area of an extremal  surface, $X$,
embedded inside the bulk spacetime \cite{RT}
\be
S_{X}={1\over 4 G_N} Area[X]
\ee  

Following an early work on $AdS_5\times S^5$ D3-brane  vacua
and  $a=3$ Lifshitz solutions
\cite{hs10}, we generalized that very approach to include all
D$p$-brane AdS vacuas  and obtained  
Lifshitz and Schr\"odinger like  solutions   
in type II A/B  string theory \cite{hs12}.\footnote{
As per our terminology,
the  `Schr\"odinger-like'
solutions we discuss in this paper will have appropriate
 conformal factors multiplying  the Schr\"odinger metrics 
 if an explicit compactification is performed, although it would not
be required in this work. 
Thus we would be discussing `conformally Schr\"odinger vacua' all along
in this work.  } 
Our main focus in this work  will  be on  
the Schr\"odinger-like  D$p$ brane solutions of \cite{hs12}. 
They exist in various dimensions with a line element
\be
ds^2_{Sch}= 
 -{\beta^2\over  z^{2a}} (dx^{+})^2 +{1\over z^2}\{-dx^{+}dx^{-}
+d x_1^2+\cdots+dx_{p-1}^2  +dz^2\},
\ee
which have a  dynamical exponent of time  as $a$. We shall mainly 
study vacua for which   $a={2\over p-5}$. 
So $a$ is  essentially negative for all 
$1<p<5$. Another special characteristic of these 
solutions is that they would involve a compact  direction, 
namely $x^-$, which is typically null. 
Due to that these type of
 classical geometries  may not be trustworthy and would require
 quantum corrections to be included. 
Although, one may choose to decompactify the null (lightcone) coordinate, 
but  a meaningful non-relativistic
Schr\"odinger group description \cite{son,bala,malda},  would
require the  lightcone coordinate  
to be rather compact. Otherwise also, even in the  noncompact cases the 
 horizon like unphysical surfaces may be present \cite{Singh:2013}.   
In  our previous work \cite{Singh:2013},
 we studied  Schr\"odinger-like 
 D$p$-brane solutions with $a={2\over p-5}$, 
but these solutions could be trusted  in 
a limited UV region only. 
The reason is that the unrestricted
IR geometry of such solutions  \cite{Singh:2013} had
 horizon-like features or a conical singularity.
Thus these  Schr\"odinger solutions with $a<0$ still
remain the least understood  D$p$-brane vacua.\footnote{We note that 
a similar type of situation arises for the Schr\"odinger vacua
 with $a>0$, namely the work  \cite{malda} for $a=2$, where 
the singularity (shrinking circle) appears in UV regime. However, 
an inclusion of black hole in $a=2$ 
solutions temporarily solves the problem.}  

In this work  we shall study AdS-bubble solutions, which 
are  well behaved geometries and  altogether avoid  IR
singularity. Remarkably, once these bubble solutions are expressed 
in  the lightcone coordinates, not only do
they mimic  Schr\"odinger-like spacetimes but  also
exhibit the same value of $a={2\over p-5}$
 as their unregulated counterpart 
in \cite{hs12}. The dynamical exponent of time 
 surely remains negative but the energy
 spectrum has no superluminosity or unrestricted blue-shifts. 
This becomes possible
 due to the presence of an in-built IR cut-off in the bubble geometries. 
Thus keeping a finite cut-off is thus essential for avoiding the singularity.
To explore these  spacetimes with $a<0$
further,  we employ a time-like T-duality \cite{Hull} and 
obtain corresponding 
time-dualized `E$p$-brane bubbles', which are solutions in
 type-II$^\ast$ superstring theory.
In  these static E$p$-brane bubbles  the dynamical exponent of time, 
however, becomes positive definite. In fact,  all E$p$ branes  have a fixed 
dynamical exponent and that is $a_{E_p}=1$.
Further,  in  E$p$-brane bubble geometries  it is rather straight forward
 to pick up  an static embedding extremal 
surface and  evaluate the  entanglement entropy.  
Particularly,  E3-brane  solutions are  used to calculate the
entropy of a strip-like subsystem in the boundary theory. To caution here,
we have literally assumed that the Ryu-Takayanagi area entropy 
functional can be suitably applied for  E$p$-branes  also, 
although these are the vacua of 
type II$^\ast$ A/B superstrings.\footnote{ A defining
feature of the type II$^\ast$ superstring action  \cite{Hull}  
is that all Ramond-Ramond  potentials have {\it negative} sign kinetic terms.
But  a complete understanding of the 
type II$^\ast$ superstrings
is still lacking in string theory. 
This work may be taken as an attempt in that direction.}

The paper is arranged as follows. In  section-2 we  review 
AdS-Bubble solutions  in lightcone coordinates
where their Schr\"odinger  spacetime
properties become quite explicit. These bubbles are smooth vacua with 
dynamical exponent $a<0$,
 without IR (conical) singularity. 
The  entanglement entropy is calculated next. In section-3 
we  employ a time-like T duality so as  
to convert these $a<0$ AdS-bubble solutions into 
 E$p$-brane bubble solutions with $a>0$. Especially we calculate the
entanglement entropy  for  the bubble E3-branes. It is found that
the entanglement entropy  of `bubble E3-branes'
 has the same functional form as it is for the `bubble D3-brane'
solutions. 
This result raises one vital question. 
Does our result imply that the field theory living on the
boundary of E3-brane bubbles has similar entanglement 
information (involving physical degrees of freedom) 
as it does for the CFT on the boundary of
  D3-brane bubbles? The answer we get
is positive. The summary is provided  in the section-4.

\section{Bubble geometries and  entropy}
We start with static  AdS-bubble or `AdS soliton'  spacetimes  
which are asymptotically
 AdS solutions in type II A/B superstring theory.
The AdS bubble solutions 
are  known to describe the low temperature 
phases in the holographic dual Yang-Mills theories at large 't Hooft
coupling. 
Typically the boundary CFT reduces to a  pure 
confining Yang-Mills theory with
 an IR cut-off  and it develops  a  mass gap \cite{witt98}. 
These  bubble D$p$-brane solutions are written as 
\bea\label{sol3h}
&&ds^2_{Bubble}=R_p^2 {r^{p-3\over2 }}\bigg[ r^{5-p}(-dt^2+fdy^2 +
d\vec x_{(p-1)}^2)  
+{ dr^2\over f r^2}  + d\Omega_{(8-p)}^2 \bigg] ,\br
&& e^\phi=(2\pi)^{2-p}g_{YM}^2({r^{7-p}\over R_p^4})^{p-3\over4}
\eea 
with appropriate electric or magnetic
flux of the Ramond-Ramond 
$F_{p+2}$ form field strength, which
 we avoid writing  explicitly as it is not  required in this work.  
The RR-flux is measured in the units of 
$N$ number of branes ($(R_p)^4\equiv {d_p g_{YM}^2 N}$). The most notations here
are the same 
as in the case $N$ Dp-branes 
\cite{itzhaki},  or see \cite{hs12}. 
But there is no supersymmetry  in the AdS bubble solutions.
We shall mainly consider the cases of D$p$-branes for $1< p <5$. 
In the above 
\be 
f(r)=(1-{r_0^{7-p}/r^{7-p}})~~~~~~~
 {\rm with}~~~~~~r_0\le r\le \infty . 
\ee
Since the radial coordinate is restricted in IR  
these describe what is commonly known as `bubble'  geometries. 
The inside of the $r=r_0$ is just empty.
As $r$ is  holographic (energy) coordinate,
the boundary  Yang-Mills theory has an effective IR cut-off  
$r=r_0$ \cite{witt98}. At length scales larger than
${1\over r_0}$ there are no correlations in the field theory due to the 
mass gap. 
However the coordinate $y$ has to be taken with right periodicity 
so that the metric  is smooth and avoids  (conical) singularity 
at $r=r_0$.  

Asymptotically, as $r\to\infty $, the solution becomes 
 conformally     $AdS_{p+2}\times S^{8-p}$,  usually with
 a running dilaton field except for a D3-brane. 
The YM theory at low temperatures can  be studied, 
in that case the Euclidean time, $\tau$ $(\equiv i t)$, will have to be  periodic. 
However the period of $\tau$ can be arbitrary in these solutions.

\subsection{Lightcone coordinates and Schr\"odinger-like bubbles}  
By introducing the lightcone coordinates ($x^\pm=t\pm y$)
 the  bubble metric \eqn{sol3h} can be expressed as
\bea\label{sol3hw}
&& ds^2=R_p^2 z^{p-3\over p-5} \bigg[ {1\over z^2}
\{ -{1-f\over 4} [(dx^{+})^2 +  (dx^{-})^2] -{1+f\over2}  dx^{-} dx^{+}
+{d\vec x_{(p-1)}^2}\br  
&&~~~~~~~~~~~~~~~~
+{4\over (5-p)^2}{dz^2\over f }\} 
+ d\Omega_{(8-p)}^2 \bigg] \br
&& e^\phi=(2\pi)^{2-p}g_{YM}^2(R_p^4 z^{2p-14\over p-5})^{3-p\over4}
\eea 
alongwith the associated $F_{+-x_1...x_{p-1}z}$ RR-flux component.
The  $z$-coordinate
has been introduced  through $r^2=z^{4\over p-5}$.
The function
\be
f(z)=1-({z\over z_0})^{2p-14\over p-5}\ee
and the  coordinate range is
$0\le z\le z_0$. From the metric \eqn{sol3hw} we see that 
the  lightcone coordinates are quite symmetrically placed, 
but both  have  time-like signatures. Thus
anyone can be tagged as the lightcone time. 
Picking $x^+$ as the time, the metric \eqn{sol3hw} can  be expressed as
\bea
\label{sol3hwwjb}
&& ds^2= 
R_p^2 z^{p-3\over p-5} \bigg[ 
-{z_0^{14-2p\over p-5}\over 4z^{4\over p-5}} (dx^{+} + U dx^{-})^2 + 
{fz_0^{2p-14\over p-5}\over  z^{4p-24\over p-5} } 
(dx^{-})^2 +{d\vec x_{(p-1)}^2 \over z^2} +{4\over (5-p)^2}{dz^2\over f z^2} + 
d\Omega_{(8-p)}^2 \bigg] 
\br && ~~~~
\equiv
R_p^2 z^{p-3\over p-5} \bigg[ 
-{\chi\over 4z^2} (dx^{+} + U dx^{-})^2 + {f\over \chi z^2 } 
(dx^{-})^2 +{d\vec x_{(p-1)}^2 \over z^2} +{4\over (5-p)^2}{dz^2\over f z^2} + 
\cdots \bigg] 
 \eea 
where  we have defined 
\be \label{lk43}
\chi(z) \equiv ({z\over z_0})^{2p-14\over p-5},~~~~~~
U(z)= {(1+f)  \over  \chi} 
\ee
for  later convenience.
Thus for the metric in \eqn{sol3hwwjb}
the effective dynamical exponent of (lightcone)  time is $a={2\over p-5}$, 
and  it is negative for  $1\le p\le 4$. 
Indeed $x^-$ coordinate  behaves as  spatial coordinate. 
In the neighborhood of $z=z_0$, by defining a new 
coordinate $u$, and since $f(z)|_{z\to z_0}\simeq u^2\to 0$,
the metric \eqn{sol3hwwjb} becomes approximately
\bea\label{sol3hwwjv}
&& ds^2\simeq  z_0^{p-3\over p-5} \bigg[ 
{1\over z_0^2} \{-
(dt)^2  + {u^2  } 
(dx^{-})^2 +{d\vec x_{(p-1)}^2 }\} +{4\over (7-p)^2}{du^2 }  + 
d\Omega_{(8-p)}^2 \bigg]  \eea 
Thus these solutions are smooth near $z=z_0$ provided we take 
$x^-\sim x^-+2\pi r^-$, with  $r^-={2 z_0\over 7-p}. $
Let us below summarize below what  have we achieved by introducing
 the lightcone coordinates in the AdS bubble metric.
\bi
\item
The   metric \eqn{sol3hwwjb} in lightcone coordinates  represents 
 conformally Schr\"odinger bubble spacetime, with 
 dynamical exponent of time   $a={2\over p-5}$. 
 These however remain smooth geometries in the IR. 
\item
Eventhough the dynamical exponent 
  is negative definite for $p\le 4$, there is no superluminosity anywhere
in the valid holographic region $z_0\ge z\ge 0$. It has got
 $\chi \le 1$ in the allowed  range. 
\item These solutions thus represent a
 resolved version of  otherwise singular Schr\"odinger-like solutions
with  same value of 
$a={2\over p-5}$,  reported in \cite{hs12, Singh:2013}. 
Also see some  details in the  Appendix.
\ei

\begin{figure}[h]
\centerline{\epsfxsize=2.5in
\epsffile{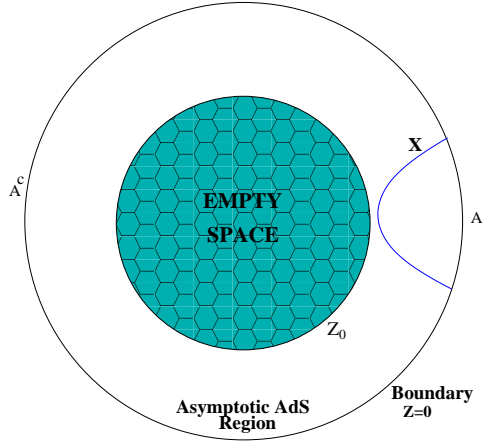} }
\caption{\label{figure4} \it The schematic drawing of a regular
Schr\"odinger/bubble spacetime when viewed in the light-cone coordinates.
Near $z\sim 0$ the spacetime becomes conformally AdS.  
The (blue) central IR region ($z>z_0$) is an empty space 
and does not exist in the geometry.}
\end{figure}

\subsection{Entanglement from bubbles}
We now obtain the entanglement entropy of a subsystem on the 
boundary of the Schr\"odinger bubble vacua \eqn{sol3hw}.
We follow  covariant  Ryu-Takayanagi 
 \cite{hubeny}, 
embedding of a strip-like  surface,
namely $x^+= y,~x^-=- y, x_1=x_1(z)$ 
inside the bulk geometry  \eqn{sol3hw}. 
The boundaries of the 
extremal bulk surface  coincide with the two ends 
of the interval $-l/2 \le x^1 \le l/2$.   
The regulated  size of the
rest of the coordinates is taken as
$0\le x^i\le l_i$, with $l_i\gg l$. We shall always have $0\le y\le 2\pi r^-$
as it is a bubble geometry, and our covariant embedding implies that on the
embedding section $x^+$ and $x^-$ will  have the same periodicity. 
Considering the  metric \eqn{sol3hwwjb},
we get the area  functional of extremal surface as 
\bea\label{schkl1saa}
 A && \equiv 
{2\pi r^- V_{p-2} \Theta_{8-p} ({L}_p)^8 \over 2G_{10}}
\int_{z_\ast}^{z_\epsilon}dz~ z^{p-9\over 5-p} \sqrt {f} 
\sqrt{{4\over (5-p)^2}{1\over f} +({\partial_z x^1})^2}
\eea  
where  
$G_{D}$ is $D$-dimensional Newton's constant, $\Theta_n$ is the
complete solid angle of  $n$-sphere, and $V_{p-2}
\equiv l_2 l_3 \cdots l_{p-1}$ is 
the  spatial volume of the subsystem. 
The new constant $L_p$ is defined as
\be
 {L}_p\equiv {(2\pi)^{p-2\over 4} \over \sqrt{g_{_{YM}}}} (R_p)^{p+1\over 4}
\ee
Note $(L_p)^2$ is an overall constant factor multiplying the metric 
\eqn{sol3hwwjb} when written in Einstein frame.
Note that the integrand in the action is
well defined in the  region, $z\le z_0$. 
In our notation $z_\epsilon\approx 0$ denotes the UV cut-off to regulate the
divergences near the boundary, 
and $z_\ast$ is the point of return
of the extremal surface  inside the bulk. 
From \eqn{schkl1saa} it follows that  
a minimal surfaces will satisfy 
\bea\label{kl3}
{dx^1\over dz}\equiv {2\over 5-p}  {({z\over z_c})^{9-p\over5-p}}{1\over
 \sqrt{f-({z\over z_c})^{18-2p\over 5-p}}} 
\eea 
The constant $z_c$  is fixed only in terms of
  the turning point   relation
\be
{f^\ast}- ({z_\ast \over z_c})^{18-2p\over 5-p}=0 
\ee
where  $f^\ast=f(z)|_{z_\ast}$.
The identification of the boundary leads to 
\be
l/2= 
{2\over 5-p}  \int^0_{z_\ast} 
{({z\over z_\ast})^{9-p\over5-p}}{\sqrt{f^\ast} \over
 \sqrt{f^2-ff^\ast ({z\over z_\ast})^{18-2p\over 5-p}}} 
\ee
which related $l$ with $z_\ast$.   
While the turning-point has the value $x^1(z_\ast)=0$. 
Evaluating it,  the  expression of the entanglement entropy
for these bubble solutions is
\be\label{kl1kv}
S_{EE}(Dp)=
{2\pi r^-  V_{p-2} ({L}_p)^p \over G_{p+2}}{1\over 5-p}
\int_{z_\ast}^{z_\epsilon}dz ~  z^{9-p\over p-5}{\sqrt{ f} 
\over\sqrt{f- f^\ast({z\over z_\ast})^{18-2p\over 5-p}}} 
\ee  
For $p=3$ the above  result gives
\be\label{kl1kvd}
S_{EE}(D3)=
{\pi r^-  l_2 ({L_3})^3 \over G_{5}}
\int_{z_\ast}^{z_\epsilon}dz ~ {1\over z^3}{ \sqrt{f} 
\over\sqrt{f- f^\ast({z\over z_\ast})^{6}}} 
\ee  
which was initially derived in \cite{takaya06}
involving $N$ D3-brane bubbles. The expression \eqn{kl1kvd}
 gives the entanglement entropy 
for a strip-like subsystem in a confining Yang-Mills  theory.

\section{{\it Time} T-duality and E$p$ branes}
  
We would like to time-dualize the AdS-bubble solutions
to obtain the dual solutions (with $a>0$) 
which can only be described as vacua of
type II$^\ast$ A(B) string theory.
However, an useful property of
the time-dual solutions would be that they admit unambiguous constant 
time hypersurfaces inside the bulk geometry, and that choice 
allows us to estimate the entanglement entropy
of the boundary theory  of E$p$-branes. 
We are
expecting that the entropy will have similar result in the time-dual cases, 
eventhough it requires for us to work in type II$^\ast$ string framework. 

For our purpose  we wish to explore the Schr\"odinger-like 
vacua \eqn{sol3hw} with $a<0$. 
We shall employ the time-like T-duality (simply TT-duality)
proposed by C. Hull \cite{Hull} for these vacua. 
Note that unlike  standard  T-duality, 
$R\to {\alpha' \over R} $, which is 
performed along an spatial circle, 
the time-like T-duality 
is performed along a (periodic) time 
direction. In  static solutions the  time is an 
isometry direction, so it will not be difficult
to make it periodic.
\footnote{ 
One can  understand this duality
  on the basis of Euclidean time as well.
Let us consider 
that `Euclidean time' has got a period,
$$
\tau \sim \tau + \beta 
 $$
then under TT-duality the `dual Euclidean time' $(\tilde\tau)$ will have
the period  
$$
\tilde \tau \sim \tilde \tau + {\alpha'\over \beta}
$$ as viewed
in the string coordinates.}
 The rules of implementing  TT-duality  are 
 discussed  in the  works \cite{Hull}. 
Using these duality maps one can generate  brane-like E$p$ solutions of the  
type II$^\ast$ A(B) theory
starting from the known D$p$-brane solutions of (ordinary) type II B(A) string
theory, and vice-versa. 

Let us consider the  AdS-bubble solutions in 
lightcone coordinates \eqn{sol3hw}.
Note that we have singled out $x^+$ as the (lightcone)
time coordinate in these solutions. We seek to make  TT-duality
along $x^+$
 and obtain the corresponding time-dual solutions in  type II$^\ast$  
string theory.
Adopting the TT duality rules given in \cite{Hull}, 
 the time-dual solutions are obtained as,
following from \eqn{sol3hwwjb} and \eqn{sol3hw}, 
\bea\label{sol4ha1}
&& d\tilde{s}^2 (E_p)=
R_p^2 z^{p-3\over p-5} \bigg[{1\over z^2}  \{ {f\over  \chi } (dx^{-})^2
+{d\vec x_{(p-1)}^2 }
+{4\over (5-p)^2}{dz^2\over f } \}
-{4z_0^{2p-14\over p-5}\over R_p^4 z^2} (d\tilde x^{+})^2 
+ d\Omega_{(8-p)}^2 \bigg] \br
&&
 e^{2\tilde\phi}= e^{2\phi}  {z^{p-7\over p-5}\over  R_p^2\chi}
=(2\pi R_p)^{4-2p}g_{YM}^2
 {z^{(p-7)(4-p)\over p-5} \chi^{-1}}
\br
&&  \tilde F_{-x_1 \cdots x_{p-1}z}= - F_{(+)-x_1 \cdots x_{p-1} z}
 ,~~~~~ \tilde B_{+-}= U(z), 
\eea 
where we have used $\tilde{}$ sign to denote the background
fields in type II$^\ast$  theory. The functions $f,~\chi,~U$ are
the same as given in \eqn{lk43}.
The $\tilde x^+$ is new (dual) time coordinate. 
The  solutions \eqn{sol4ha1} 
are generally  recognized as E$p$-branes\footnote{ 
An E$p$-brane has  $p$-dimensional
Euclidean world-volume and appears as a solution  
in type II$^\ast$ A/B string theories \cite{Hull}. Note, for all E$p$-branes,
 especially the  time coordinate counts as
one of the $(10-p)$ transverse (Dirichlet) directions, 
whose all  world-volume (Neumann) coordinates are otherwise Euclidean. 
An E0-brane is localized in 10-dimensional spacetime.}. 
The   coordinates patch $ (x_1,\cdots, x_{p-1},x^-)$ in  eq.\eqn{sol4ha1} 
defines a $p$-dimensional
Euclidean world-volume of an E$p$-brane.
In the present case the 
E$p$-branes also have nontrivial $B_{\mu\nu}$ field, which
 implies  the presence of a fundamental string,  stretched along  $x^-$ 
direction along the  world-volume. 
Note that the  time coordinate $\tilde x^+$ should  be treated 
as  the Dirichlet coordinate and is obviously
transverse to the E$p$-brane world-volume.
The E$p$-branes are fundamentally
charged under $\tilde A_{(p)}$,  the $p$-form  RR potential.
It is  remarkable that
the dynamical exponent of  time is simply $a=1$ for all the above
E$p$-branes in \eqn{sol4ha1},
while one spatial coordinate, namely $x^-$, has an effective dynamical exponent
${2p-12\over p-5}$ (note $\chi\equiv ({z\over z_0})^{2p-14\over p-5}$). 
That generates 
asymmetric scaling which may be alluded to the presence 
of fundamental string in the E$p$ solutions.

\subsection{`D3 with $a=-1$' to `E3 with $a=1$'}

For the simplification purpose,
we shall only discuss the 
D3-brane backgrounds here, but   the procedure
can be repeated for all other D$p$ brane cases also. 
The Schr\"odinger-type $(a=-1)$  D3-brane bubble vacua
in type IIB string theory, read from eq.\eqn{sol3hw}, is  
\bea\label{sol4hw}
&& ds^2_{D3}=
(R_3)^2  \bigg[  -{1\over 4 } {z^2\over z_0^4} [dx^{+} +  
(1+f){z_0^4\over z^4}dx^{-}]^2 
+  {z_0^4\over z^6}{f} (dx^{-})^2
+{d\vec x_{(2)}^2\over z^2} 
+{dz^2\over f z^2} + d\Omega_{(5)}^2 \bigg] \br
&& e^{\phi_b}={g_{YM}^2\over 2\pi},~ ~~ F_{(5)}^b=4 R_3^4(1+\star)\omega_5\ .
 \eea 
where $F_{(5)}$ is self-dual, $\omega_5$ is the unit volume
element over unit $S^5$,
and $f(z)=1-{z^4\over z_0^4}$.
Note that $x^+$ is  the time coordinate, say  with its (Euclidean) 
 period $\beta$. 
Then the  period of Euclidean time,  
fixed by a measurement say at $z=z_0$ inside the bulk, 
will be   ${\beta R_3 \over 2 z_0 }$. By making  TT-duality
along $x^+$ direction
 and we shall obtain the corresponding  time-dual solution of  type II$^\ast$A 
theory.
From \eqn{sol4ha1} we obtain the time dual solution as
\bea\label{sol4ha}
&& d\tilde{s}^2_{E3}=
R_3^2  \bigg[   {z_0^4\over z^6}{f} (dx^{-})^2
+{d\vec x_{(2)}^2\over z^2}  
+{dz^2\over f z^2}
 - {4z_0^4\over (R_3)^4 z^2} (d\tilde x^{+})^2 
+ d\Omega_{(5)}^2 \bigg] \br
&&
 e^{2\tilde \phi_a}=
 e^{2 \phi_b}
{4 z_0^4\over R_3^2  z^2},
~~~~~\tilde B_{+-}= (1+f){z_0^4\over z^4}, 
\br
&&
\tilde F^{a}_{-x_1 x_2 z}=- F^{b}_{(+)-x_1 x_2 z}
\eea 
We have used the suffix ${a(b)}$ in order to distinguish
type II$^\ast$A and type IIB solutions. 
The above solution \eqn{sol4ha} is an  E3-brane bubble.
Three  coordinates $(x^-,x_1,x_2)$ in  \eqn{sol4ha} 
define 3-dimensional
Euclidean world-volume of  E3-brane.
Note that the dual time coordinate $\tilde x^+$ will be treated 
as one of  the Dirichlet coordinates and is obviously
transverse to the E3-brane world-volume. 
Also note that  the
 period of dual (Euclidean) time, would be taken 
$\tilde\beta= {1\over\beta}$. 
Correspondingly the period of dual time 
inside the bulk, measured at $z=z_0$, becomes ${2 z_0\over R_3}\tilde\beta$. 
The E3-branes are fundamentally
charged with $F_{(4)}$  RR field strength. 

Since we have $x^-\sim x^- + 2\pi r^-$,
the solutions \eqn{sol4ha}
are smooth near $z=z_0$ 
with the radius  $r^-= z_0/2$. While
near the boundary, as $z \to 0$, the E3 solution \eqn{sol4ha}
simply becomes
\bea\label{sol4has}
&& ds^2_{E3}\simeq
R_3^2  \bigg[   {z_0^4\over z^6} (dx^{-})^2
+{d\vec x_{(2)}^2\over z^2}  
+{dz^2\over  z^2}
 - {4z_0^4\over R_3^4 z^2} (d \tilde x^{+})^2 
+ d\Omega_{(5)}^2 \bigg],\br
&&  
 e^{2\phi_a}={g_{YM}^2\over 2\pi} {4 z_0^4\over R_3^2 z^2}, 
~~~~~B_{+-}\simeq 2{z_0^4\over z^4}, \br
&&
F^{(a)}_{-x_1 x_2 z}\simeq F^{(b)}_{(+)-x_1 x_2 z}
\eea 
which is an anisotropic spacetime. 
This asymptotic form of metric \eqn{sol4has} 
has an explicit asymmetric scaling property:
\bea
z\to \zeta z, ~~ 
\tilde x^+\to \zeta \tilde x^+,~~
x^-\to \zeta^3 x^-,~~
x^1\to \zeta x^1,~~
x^2\to \zeta x^2
\eea
Note the  scaling however requires  $r^-\to \zeta^3 r^-$,
so the compactification radius ought to change alongwith  the scaling. 
It  thus is a  source of  scaling violation. It essentially  results in 
hyperscaling violation in  lower dimenional action. 
Overall the world-volume  coordinate $x^-$ 
scales differently as compared to the rest, namely $x^1$ and $x^2$. 
Also the  presence of  $B_{+-}$ 
breaks the $SO(3)$  invariance down to rotation in $x_1-x_2$
 plane on E3-brane
world-volume. However, the dynamical exponent of 
time is  $a=1$, so it
behaves like  a relativistic field theory.  
Actually above  E3 branes are   delocalized 
along the time $\tilde x^+$ direction, so the time  is not essentially a
decoupled direction unlike the 5-sphere in the geometry \eqn{sol4ha}.
The field theory thus lives
on the boundary of  5-dimensional spacetime 
$$\bigg[ {z_0^4\over z^6}f (dx^{-})^2
+{d\vec x_{(2)}^2\over z^2}  
+{dz^2\over f z^2}
 - {4z_0^4\over R_3^4 z^2} (d\tilde x^{+})^2 \bigg]$$
 In the
next section our aim is to determine the entanglement entropy of 
the boundary theory.   
Also note that all constant time surfaces in \eqn{sol4ha} 
are smooth spatial geometries.

\subsection{The entanglement entropy of E3-brane bubbles}

 Here we shall first assume 
that the Ryu-Takayanagi proposal works also for the case of E$p$-brane
solutions of type II$^\ast$ string
theories. The reason for this belief 
is that the entanglement entropy is calculated 
 geometrically  as an area of an extremal entanglement surface.
There exist  unambiguous spatial slices described by 
$x^+= {\rm constant}$ surfaces in \eqn{sol4ha}.  
  So we  look for a  static (constant $x^+$)   
embedding of a strip-like 3D spatial surface, 
namely $(x^-, ~x^1(z),~ x^2)$ inside the bulk geometry 
described by   \eqn{sol4ha}. 
The end points of the 
3D surface  coincide with the boundaries of the strip $-l/2\le x^1\le l/2$.   
The  size of the
rest of spatial coordinates is taken as
$0\le x^-\le 2\pi r^-,~0\le x^2\le l_2$, with the length $l_2\gg l$.
Considering the  metric in \eqn{sol4ha},
 we determine the area   functional of entanglement surface 
\bea\label{schkl1s}
 \tilde A && \equiv 
{\pi  r^-l_2 \Theta_5 R_3^8\over \tilde G_{10}}
\int_{z_\ast}^{z_\epsilon}dz e^{-2\phi_a} { z_0^2\over z^5 }\sqrt {f} 
\sqrt{{1\over  f} +({d x^1\over d z})^2}
\br && =
{\pi  r^-l_2 \Theta_5 R_3^8\over \tilde G_{10}}
{e^{-2\phi_b} R_3^2\over 4 z_0^2}
\int_{z_\ast}^{z_\epsilon}dz  { \sqrt{\chi}z_0^2\over z^5 }\sqrt {f} 
\sqrt{{1\over  f} +({d x^1\over d z})^2}
\br && =
{\pi r^- l_2 L_3^3  \over \tilde G_5 }{ R_3^2\over 4 z_0^2}
\int_{z_\ast}^{z_\epsilon} dz  {\sqrt {f} \over z^3 } 
\sqrt{{1\over  f} +({\partial_z x^1})^2}
\eea  
where ${1\over \tilde G_5}={\Theta_5 L_3^5\over \tilde G_{10}}$. 
Note that $z_\epsilon\approx 0$ denotes the UV cut-off
and $z_\ast$ is  the turning point
of the extremal surface inside the bulk. 
From \eqn{schkl1s} we can determine that  
the minimal surface must satisfy the  equation
\bea\label{klv3}
{dx^1\over dz}&=&  {({z\over z_\ast})^{3}}{
\sqrt{f^\ast}\over \sqrt{f^2-ff^\ast({z\over z_\ast})^6}} 
\eea 
where  we have denoted $f^\ast=f(z_\ast)$.
The identification of the boundary is $x^1(z_\epsilon)=l/2$, 
   while the turning-point itself has the mid-point value $x^1(z_\ast)=0$. 
Note that we shall have to take $z_\ast \le z_0$ always.
Thus we get the  expression of the entanglement entropy
for the YM theory living on the boundary of the E3 solutions
\bea\label{schkl1ka}
&&\tilde S_{EE}(E3)=
{\pi  r^- l_2 (L_3)^3\over \tilde G_5}{ R_3^2 \over 4 z_0^2}
\int_{z_\ast}^{z_\epsilon}dz {1\over  z^3}{ \sqrt{f} 
\over\sqrt{{f}- f^\ast({z\over z_\ast})^{6}}} 
 \eea  
This is a complete  expression for the entanglement entropy
of the boundary  theory
of E3-brane bubbles, where the boundary subsystem is a flat strip of width $l$
along $x^1$ and covers the whole of  $x^-$  and $x^2$. It is interesting that,
apart from  numerical factors outside the integral in \eqn{schkl1ka},  
the  integral expression of the entanglement entropy is the same as it 
is found in the case of $AdS_5$-bubble  vacua 
involving D3-branes in \eqn{kl1kvd}. 

Clearly the two entropy  expressions \eqn{schkl1ka}  and \eqn{kl1kvd}
look similar, but differ  in  respective  numerical factors multiplying
them, which we do not expect 
to be same given that we have evaluated entropy using constant time slices. 
But the two integrals in them remain essentially the same.
It is a well known fact that static $AdS$-bubble vacua
have lowest entropy amongst the asymptotically $AdS$ vacua with same symmetry
\cite{witt98,takaya06}. 
Taking this fact as our guiding principle, we expect that  the 
measurements of entropy  for
 E3-bubbles should also not deviate from this. The integral
form of expression in \eqn{schkl1ka} essentially favours the proposal 
that the entanglement entropy of  E3-brane bubbles is also the lowest amongst
all E3-brane vacua having  same asymptotic symmetry.

\section{Summary}

We  have studied     
D$p$-brane bubble solutions having  conformally AdS asymptotic 
geometries. 
Once these are expressed in lightcone coordinates the solutions mimic 
Schr\"odinger-like geometries   with  dynamical exponent of time  
given by $a={2\over p-5}$. Thus $a$ is essentially negative for $p<5$. 
These solutions are smooth geometries in the  IR and have a cut-off. Thus the 
entanglement entropy is well defined.  
To convert these solutions into the solutions with exponent $a>0$,
we  employed  time-like T-duality. 
The time-like duality of course gives rise to   E$p$-brane solutions 
with $a>0$. In these E$p$-brane bubble geometries finding  
static  embedding surfaces
is rather straight forward. Particularly, we have  
 estimated the entanglement entropy of a strip-like subsystem on 
the boundary of  E3-brane bubble solution. It is found that, 
barring an overall numerical factor, the entanglement entropy
has the same functional form as it were in the case  of   D3-bubble
solutions \cite{takaya06}. 
This new result involving Ep-branes is 
interesting  but it leaves some unanswered questions. First, why 
the entanglement entropy has to have the same functional form for 
 E3-bubbles and the $AdS_5$-bubble cases. Does it imply that the
entropy of E3-branes counts the right physical degrees
of freedom, eventhough the Lagrangian formulation 
of type II$^\ast$ superstrings 
are not so well understood, as these theories come with
 negative sign kinetic terms in the   RR-sector.
We may note that E$3$ branes preserve supersymmetry but E3-bubbles do not. 
Secondly, does the  entropy (except an overall constant) 
mean that we could
trust type II$^\ast$ string theories in  
asymptotically AdS E$p$-brane background, like the ones presented here.  
Our minimal entropy result for E3-brane bubbles is encouraging and 
will have some important implications on these issues. 
We hope to report on some of them	
in the next work \cite{hs15}.

\vskip.5cm
\noindent{\it Acknowledgments:} 
I wish to thank the organizers of the 
`Workshop on Black Hole Information Paradox' at HRI Allahabad, 
 for the hospitality during which
this work got partly done. 
I also thank  Sandip Trivedi for an useful discussion.

\vskip.5cm
   
\appendix{

\section{ Conformally Schr\"odinger spacetimes with $a<0$
and Entanglement}

A class of  conformally
Schr\"odinger  spacetimes with negative dynamical exponents of time
were  reported in \cite{hs12}. These could be generated 
by taking  double limits
of the boosted bubble solutions or by  Wick rotations
of the corresponding Lifshitz type solutions. 
These solutions are  (for $p\ne 5$)
\bea\label{sol3h4}
&&ds^2_{Sch}=R_p^2 z^{p-3\over p-5} \bigg[ 
\{ -{1\over 4z^2}({z\over z_s})^{2p-14\over p-5} (dx^{+})^2 +{-dx^{+}dx^{-}
+d\vec x_{(p-1)}^2\over z^2}  
+{4\over (5-p)^2}{dz^2\over  z^2}\}  + d\Omega_{(8-p)}^2 \bigg] \br
&& e^\phi=(2\pi)^{2-p}g_{YM}^2({R_p^4 z^{2(p-7)\over p-5}})^{3-p\over4}
\eea 
with  $(p+2)$ form RR-flux. The $z_s$ is an intermediate scale  in the IR regime.
One can see that there is an asymmetric scaling  involving 
the coordinates as \cite{hs12}
\be\label{bh3}
z\to \xi z , ~~~~
x^{-}\to \xi^{2-a}  x^{-},~~~~
x^{+}\to \xi^{a} x^{+},~~~\vec x\to \xi \vec x
\ee
with dynamical exponent 
\be
a={2\over p-5},
\ee 
under which 10-dimensional dilaton and the string metric conformally
scale as
\bea\label{jhi1}
g_{MN}\to \xi^{p-3\over p-5} g_{MN},
~~~e^\phi\to \xi^{(7-p)(p-3)\over2(p-5)}e^\phi
\eea
The latter equation is  the standard Weyl (conformal) scaling
behavior of the near-horizon D$p$-brane vacuas \cite{itzhaki}
and it remains unchanged. Notice that,
since $x^-$  is to be taken compact
for a nonrelativistic (Schr\"odinger group) realisation of the CFT,
the scaling, namely $x^{-}\to \xi^{2-a}  x^{-}$ above,
involves jumps in the compactification radius. Thus the rescaling will
take us from one compactification radius to another, but preserving 
the Weyl scaling \eqn{jhi1} of the metric. 
The solutions \eqn{sol3h4}  are  invariant 
under space and time translations and  rotations in the 
Euclidean patch $\vec{x}_{(p-1)}$.

It can be  noted that the dynamical exponent 
$a$ is negative for most of
the interesting cases of $p=2,3,4$ branes in \eqn{sol3h4}.    
But there are some problems in interpreting these naive Schr\"odinger 
vacua.
\bi
\item The $a<0$  vacua would
 give rise to the spectrum which has unrestricted  blue-shift in the IR 
($g_{++}$  blows up in IR). 
\item Once 
 $x^-$ is  compactified, it leads to 
 an existence of a conical singularity \cite{Singh:2013}.  
\item Even if $x^-$ is noncompact, there would exist  `horizon' 
like surfaces in the IR region ($z \ge z_s$)
 \cite{Singh:2013}. Thus for  solutions \eqn{sol3h4} 
things start getting worse near 
$z\sim z_s$. 
\ei
All  above issues are very much interrelated! We try to 
resolve them by implementing minimal changes in the solutions
in this work and have employed  solutions with cut-off.  

\section{ Entanglement Entropy}

Let us also discuss here the entanglement entropy of a subsystem on the 
boundary of the Schr\"odinger  vacua \eqn{sol3h4}, 
at least in the UV region where we can make some definite conclusions.
We look for a  covariant  
embedding of a strip-like  surface, 
namely $$x^+=2y,~x^-=-y/2, ~ x^1=x^1(z)$$ 
inside the bulk geometry \eqn{sol3h4}. 
The boundaries of the 
extremal bulk surface do coincide with the boundaries 
of the strip having the width $-l/2\le x^1\le l/2$.   
The  volume of the
rest of the coordinates is taken as
$0\le x^i\le l_i$, with $l_i\gg l$, and we shall take $0\le y\le 2\pi R^-$.
Considering the  metric  \eqn{sol3h4},
we determine the area  functional of entanglement surface as 
\bea\label{schkl1sa}
 A && \sim 
{2\pi R^-V_{p-2} \Theta_{8-p}L_p^8 \over 2G_{10}}
\int_{z_\ast}^{z_\epsilon}dz~ z^{p-9\over 5-p} \sqrt {K} 
\sqrt{{4\over(5-p)^2} +({\partial_z x^1})^2}
\eea  
where $G_{10}$ is 10-dimensional Newton's constant, $V_{p-2}
=l_2 l_3 \cdots l_{p-1}$ is 
the total volume, and the function
$$K(z)=1-({z\over z_s})^{14-2p\over 5-p}.$$ 
Note that
the integrand of the action
becomes undefined in the   region $z\ge z_s$, 
 where $K$ becomes negative.
This is also an indication of the fact that the bulk geometry has  
unphysical region \cite{Singh:2013}, 
where the superluminal (blue-shift) effects  become prominent in IR.
In our notation $z_\epsilon\approx 0$ denotes the UV cut-off to regulate the
divergences near the boundary, 
and $z_\ast$ is the turning point
of the extremal surface. 
From \eqn{schkl1sa} we can determine that  
the minimal surfaces must satisfy the  equation
\bea\label{2kl3}
{dx^1\over dz}&\equiv & {2\over 5-p} {({z\over z_c})^{9-p\over5-p}}{1\over
 \sqrt{K-({z\over z_c})^{18-2p\over 5-p}}} 
\eea 
where $z_c$ is a constant and is fixed by  the turning point ($z=z_\ast$)
relation
\be
{K^\ast}- ({z_\ast \over z_c})^{18-2p\over 5-p}=0 
\ee
We have denoted $K^\ast=K(z)|_{z=_\ast}$.
The identification of the boundaries gives $x^1(z_\epsilon)=l/2$,
   while the turning-point has the mid-point value $x^1(z_\ast)=0$. 

Evaluating it, we get the  expression of the entanglement entropy
for the  solutions \eqn{sol3h4}
\be\label{schkl1k}
S_{EE}\sim 
{\pi R^-V_{p-2} (L_p)^p \over G_{p+2}}
\int_{z_\ast}^{z_\epsilon}dz ~  z^{9-p\over p-5}{ K 
\over\sqrt{K- K^\ast({z\over z_\ast})^{18-2p\over 5-p}}} 
\ee  
Given that $K=1-({z\over z_s})^{14-2p\over 5-p}$,  the above
result is the same as the expression obtained  in \cite{Singh:2013}
by doing an explicit compactification along the lightcone of 
Schr\"odinger vacua. But note that
this formula  can only be trusted for the small size subsystems 
for which $z_\ast \ll z_s$, due to the 
reasons of unregulated IR bulk region. But the generic nature of the
expression of $K$ is such that, it results in reduced entanglement,
when compared to the case with $(z_s)^{-1}=0$ (pure conformally AdS cases). 
}

\end{document}